\documentclass[twocolumn, preprintnumbers, aps]{revtex4}
\usepackage{pstricks,graphicx,epsfig,color,amssymb,amsmath,amscd}

\newcommand{\bi}{\bibitem}
\newcommand{\be}{\begin{eqnarray}}
\newcommand{\ee}{\end{eqnarray}}
\newcommand{\rar}{\rightarrow}
\newcommand{\Rar}{\Rightarrow}

\usepackage{latexsym}

\begin{document}

\title{Gravitational production of KK states}

\author{Cosimo Bambi$^{\rm 1}$}
\author{Federico R. Urban$^{\rm 2}$}

\affiliation{$^{\rm 1}$Department of Physics and Astronomy, Wayne State University, Detroit, MI 48201, USA\\
$^{\rm 2}$ Department of Physics and Astronomy, University of British Columbia, Vancouver, BC, V6T 1Z1, Canada}

\date{\today}

\preprint{WSU-HEP-0810}

\begin{abstract}
Gravitational particle production in the context of braneworlds is considered from a phenomenological point of view. The production of KK modes for bulk fields is discussed and their abundances computed. The results have been applied to some specific fields such as gravitinos and axions, and their cosmologies have been outlined, exemplifying the constraints on the properties of the extra dimensional model in each case.
\end{abstract}

\maketitle

\section{Introduction}

Gravitational particle production in an ordinarily expanding Friedman-Robertson-Walker (FRW) Universe dominated by dust-like or radiative matter is known to be a very poorly efficient particle creation mechanism~\cite{book}. Indeed, if one leaves the conformal anomaly exception aside, despite being incessantly operative during the whole evolution of the Universe, the abundances of non-conformally coupled particles is totally negligible. This is due entirely to the weakness of the coupling of particles to gravity, combined to the steadiness of the evolution of the Cosmos. In order for this particular mechanism to be efficient one needs to look for abrupt transitions in the dynamics of the FRW scale factor, such as that believed to have taken place during inflation, and, more importantly, when the de Sitter phase drew to an end, repopulating the Universe.

However, as it has been recently pointed out in ref.~\cite{noi1}, if the four dimensional FRW Universe were to be embedded into a higher dimensional spacetime, such as in~\cite{add-rs}, then even dust-like or radiation dominated FRW Universes would be able to inject gravitationally produced particles into the plasma in a sizeable way. This is easily understood as a consequence of the possibility that the actual scale of - 5D - gravitational interaction be tuned to much lower values, thereby drastically enhancing the coupling with matter. Hence, for the same reason one expects to see signatures of (collections of) Kaluza-Klein (KK) gravitons in TeV colliders~\cite{lhc}, in the most optimistic scenarios, it is expected that expanding cosmological backgrounds should turn out to be phenomenologically relevant as well.

This new feature of brane cosmologies shows up when one derives the effective Friedman equation for a 4D brane immersed into a 5D spacetime. In this case it is known that the evolution of the Universe changes at high energies, allowing for a new term in the Friedman equation for the Hubble parameter, which in general reads~\cite{cosmo}
\be\label{friedman}
H^2 = \frac{\rho}{3 M_4^2} \left(1 + \frac{\rho}{2 \lambda}\right) \, ,
\ee
where $\rho$ is the energy density of ordinary matter on the brane, $M_4 \simeq 2.4 \cdot 10^{15}$~TeV the 4D reduced Planck mass,
\be
\lambda = \frac{6 M_5^6}{M_4^2}
\ee
is the brane tension and $M_5$ the effective gravity - reduced - scale of the five dimensional theory. The transition temperature $T_t$ is the temperature at which the evolution of the Universe switches from braneworld regime to standard one and, if the Universe is radiation dominated, it is given by
\be
T_t = 2 \left(\frac{360}{\pi^2 g_*}\right)^{1/4} \left(\frac{M_5^3}{M_4}\right)^{1/2} \, ,
\ee
where $g_*$ counts the relativistic degrees of freedom tightly coupled with the thermal plasma.

Therefore, if the temperature of the plasma is higher than the transition temperature, the Universe would expand according to $H = \rho/6M_5^3$. One can immediately see the effect of the 5D embedding: at high energies the expansion is regulated by the 5D Planck mass rather than the low energy 4D coupling.

At this point it is useful to review the basic results found in~\cite{noi1}, which provide the basis for the extensions presented hereafter. In the following formulas everything will be expressed in TeV units, unless explicitly stated.

As already specified, for the time being the focus will be on fields non-conformally coupled to gravity, whose gravitational `charge' is their mass $m$. Gravitational particle production in a time-variable metric is in general efficient as long as the Hubble parameter $H$ is much larger than the particle mass, that is $H \gg m$. In this regime it can be shown that the number density of created particle is constant~\cite{mms}
\be\label{number}
n = \frac{m^3}{24 \pi^2} \, ,
\ee
whereas for $H \ll m$ particle creation is negligible and $n$ decreases as $1/a^3$, where $a$ is the FRW cosmological scale factor. Of course this result is slightly model-dependent, i.e.\ it depends on the specific law linking expansion rate to time, but numerical results confirm that it is accurate within an order of magnitude for any relevant power law expansion of the scale factor, $a(t) \propto t^q$. Consequently, for the sake of simplicity, for $H > m$, the particle number density will be given by eq.~(\ref{number}), and particle production is taken to cease instantaneously as soon as $H = m$. Mention to the actual numerical result will be given throughout the paper.

Within this approximation, the freezing-out of gravitational production corresponds to a temperature around
\be\label{freeze-out}
T_f^4 \simeq \frac{18}{g_*} \, m \, M_5^3 \, .
\ee

Were this species a stable one and very weakly interacting with ordinary matter, so that it would not thermalise or decay as the Universe cools down, then its relic density today would contribute with a fraction of
\be\label{abund-general}
\Omega_X h^2 \simeq 8 \cdot 10^7 \, \frac{m^{13/4}}{M_5^{9/4}} \, ,
\ee
to the total energy density of the Universe, see~\cite{noi1}, where $h$ is the present Hubble parameter in units of 100~km~s$^{-1}$~Mpc$^{-1}$. Eq.~(\ref{abund-general}) is derived assuming that at the temperature $T_f$ there are about $g_* \simeq 100$ relativistic degrees of freedom, which implies also a dilution factor of 0.04, and is valid only if the conditions
\be
T_f \gtrsim T_t &\Rar& M_5 \lesssim 6 \cdot 10^9 \, m^{1/3} \, ,\label{condition1} \\
T_f \lesssim M_5 &\Rar& M_5 \gtrsim 0.2 \, m \, ,\label{condition2}
\ee
are fulfilled, as required by consistency.

This is the result which was used in~\cite{noi1} to assess the importance of such mechanism in, e.g., the generation of sterile viable Dark Matter (DM), baryogenesis scenarios at low temperature (see also~\cite{baryogenesis}), or its potential dangerousness, had the produced particles been equipped with a decay channel timed after the synthesis of the light elements (Big Bang Nucleosynthesis, or BBN), as for instance for the gravitino case.

\section{Kaluza-Klein towers}

The situation hitherto described applies to brane-trapped fields, which therefore depend only on four coordinates, and have by definition only a zero 4D mode. However, especially in the gravitino case, it is likely that fields have access to the full dimensionality of the spacetime, and that, as long as the extra dimension is compact, from a 4D observer they would be seen as a tower of KK states. Such a configuration was indeed studied in connection with the thermal production of KK gravitinos for a number of supersymmetric (SUSY) extra dimensional models~\cite{noi2}.

The aim of this section is to extend the analysis given in~\cite{noi1} to the full tower of KK states, whereas the following ones will review its implications for the phenomenology of extra dimensions, and its constraining power applied to specific fields, one at a time placed in the higher dimensional spacetime. A general, theoretically oriented discussion about gravitational particle production for 5D fields can be found in ref.~\cite{indiani}.

\subsection*{Flat fifth dimension}

In this first scenario the extra dimension will be taken to be flat. In general the mass of each KK mode is given approximately by $m_n = m_0 + n/R$, where $R^{-1} = 2\pi M_5^3 / M_4^2$ is the inverse of the size of one compact extra dimension. This expression of course is not universally true, since an effective potential for the 5D field may appear upon dimensional reduction, and therefore contribute to the effective mass of the KK mode. Nevertheless, in general it is expected that at high $n$ this parametrisation provides a good approximation, as it will be clarified in the worked examples considered below.

Having obtained an explicit formula for the mass, one can readily write down the abundance associated to each KK mode. In the limit for which the zero mode is light compared to $n/R$, which again is taken in light of the particular models to be dealt with, the fraction of energy density and the yield variable turn out to be
\be
\Omega_X^n h^2 &\simeq& 3 \cdot 10^{-90} n^{13/4} \, M_5^{15/2} \, , \label{omegaNflat} \\
Y_X^n &\simeq& 1.3 \cdot 10^{-71} n^{9/4} \, M_5^{9/2} \, . \label{yNflat}
\ee
Clearly, eq.~(\ref{omegaNflat}) holds only for stable particles or for particles whose lifetime exceeds the age of the Universe. $Y_X^n = n_X^n / s$, where $s$ is the total entropy density of the Universe, is a conserved quantity for species which are decoupled from the thermal plasma and have not yet decayed. We notice that we are still assuming a Universe dominated by relativistic matter, where the amount of particles produced by the time-varying metric is small when this generating mechanism is efficient. These two quantities are the most useful in obtaining constraints on the parameters of the theory, and will be extensively employed.

It is immediate to rewrite the consistency conditions~(\ref{condition1}, \ref{condition2}) for this specific setup: they are
\be
T_f \gtrsim T_t &\Rar& n \gtrsim 5 \, ,\\
T_f \lesssim M_5 &\Rar& n \, M_5^2 \lesssim 5 \cdot 10^{30} \, .
\ee
What the first line means is that the results~(\ref{omegaNflat}, \ref{yNflat}) do not hold for the first handful of modes, as the quantities $\Omega_X^n$ and $Y_X^n$ have been deduced assuming $H = \rho/6 M_5^3$, while for small $n$ we would find $T_f < T_t$. For these light modes, the gravitational production stops at the transition temperature $T_t$ or, if very light, during the standard expansion, thereby rendering their abundances negligible. However, it should be stressed that is the case only if the zero mode mass is small compared to the mass gap $\sim 1/R$ between the modes. Indeed, in that situation a heavy zero mode would be copiously produced, and so would be each KK mode from the very bottom of the tower. Anyway, since such a configuration would not significantly alter the results presented below, this possibility is not discussed further.

Moreover, we stress again that our discussion is reliable for large $n$'s, while the zero and the first few modes would require a separated treatment, which depends on the exact model under investigation. For example, even a massless zero mode may be gravitationally produced, because in a more general setup the 5D metric 
\be
ds^2 = e^{2 \sigma(z)} a^2(\eta) \left[ d\eta^2 - dx_1^2 -dx_2^2 -dx_3^2 \right] - dz^2
\ee
is not conformally flat (for more details, see the discussion in sec.~8 of ref.~\cite{indiani}). In a radiation dominated Universe, that is the case of interest in the present work, there is an accidental recovery of conformal invariance, hence, no gravitational production of the zero mode via time variation of the metric (as long as the conformal anomaly is not considered, see appendix).

In general each mode will have its own interaction strength(s) (if any) with Standard Model (SM) fields, and the history of each mode would depend on the way they decay. The same statement applies to the impact on cosmological observables such as DM energy density, light elements abundances, and so on. The main discriminator among these different effects is the lifetime of the particle, which, once the zero mode's decay properties have been specified, depends almost solely on the mass of each higher mode. It is thus convenient to split the KK tower into various bands to be taken into account separately.

Once the interesting lower and upper limits have been identified, one is in a position to scrutiny the number of modes in that given range, and investigate their total impact on a particular observable. Moreover, it is reasonable to expect that only large number of states for each band could provide significantly different constraints compared to an ordinary 4D particle with mass within the given band. This last consideration, combined with the fact that the zero mode is supposed to be light, implies that the highest KK mode $N$ in the band will satisfy $N \gg N_0$, with $N_0$ the lightest mode in the same band. Then it is straightforward to obtain the overall density fraction and yield variable for each band by just summing over the modes and then discarding the (smaller) contribution of the lowest state, the result of which being
\be
\Omega_X h^2 \equiv \sum_n \Omega_X^n h^2 &\simeq& 7 \cdot 10^{-91} N^{17/4} \, M_5^{15/2} \, , \label{omegaTOTflat}\\
Y_X \equiv \sum_n Y_X^n &\simeq& 4 \cdot 10^{-72} N^{13/4} \, M_5^{9/2} \, . \label{yTOTflat}
\ee
We note in passing that the actual summation approaches the numerical results (\ref{omegaTOTflat}) and (\ref{yTOTflat}) --obtained by integration, for a number of modes $N \gtrsim 20$, beyond which the precision is better than 10\%.

\subsection*{Warped fifth dimension}

If the extra dimension is warped, the structure of the KK tower turns out to be different. In the case of two branes, typically one finds that the mass of the KK mode $n$ is
\be
m_n = m_0 + k x_n e^{-\pi R k} \, ,
\ee
where $k = (1 - \exp (-2\pi k R)) M_5^3 / M_4^2$ is the $AdS_5$ curvature, $x_n$ is the $n$-th root of the first order Bessel function $J_1$ and $\pi R$ is the size of the orbifold. The mass splitting is $\Delta m \simeq 3 k \exp(- \pi R k)$, and for later convenience we define the function $F(kR) \equiv (1 - \exp (-2\pi k R)) / \exp (\pi k R)$.

The Friedman equation is basically the same of the case with a flat extra dimension and we can still use eq.~(\ref{friedman}). It is thus fairly straightforward to repeat the steps undertaken for the flat case, with results
\be
\Omega_X^n h^2 &\simeq& 3 \cdot 10^{-91} \, n^{13/4} F^{13/4} M_5^{15/2} \, , \label{omegaNwarped}\\
Y_X^n &\simeq& 2 \cdot 10^{-72} \, n^{9/4} F^{9/4} M_5^{9/2} \label{yNwarped} \, ,
\ee
in place of eqs.~(\ref{omegaNflat}, \ref{yNflat}), and
\be
\Omega_X^n h^2 &\simeq& 7 \cdot 10^{-92} \, N^{17/4} F^{13/4} M_5^{15/2} \label{omegaTOTwarped} \, , \\
Y_X^n &\simeq& 8 \cdot 10^{-73} \, N^{13/4} F^{9/4} M_5^{9/2} \, . \label{yTOTwarped}
\ee
which now replace eqs.~(\ref{omegaTOTflat}, \ref{yTOTflat}).

It is now easy to see why the scenario is more or less unchanged in this picture, unless one employs very small values for $F(kR)$. Indeed this function is always smaller than $\sim 0.4$, for which value the abundances and consistency constraints turn out to be very similar to those previously obtained for a flat extra dimension. If one instead chooses to work with much smaller $F(kR)$, such as for $kR = 11$, then the mass splitting becomes extremely tiny, unless the five dimensional Planck mass is pushed all the way up to the 4D $M_4$. Therefore, the mass gap drops by a factor of $2/F$, and the abundances of produced particles increase by a huge factor $1/F$ (this can be seen explicitly once the expressions for $N$ --see section~\ref{app}, is plugged in~(\ref{omegaTOTwarped}) and~(\ref{yTOTwarped})). The compensation in $M_5$ is proportional to $F^{-4/21}$ which, again for $kR = 11$, is around 700, although in highly warped models it is customary to safely (as far as this mechanism is concerned) take $M_5 \simeq M_4$.

We would further like to stress that all our results are based on the validity of eq.~(\ref{friedman}). Actually, eq.~(\ref{friedman}) is deduced from the 5D Einstein equations in the case ${\it i)}$ there is no flux along the fifth dimension and ${\it ii)}$ the bulk energy density is constant, which is the case for the standard scenarios where particles are confined to the brane and the energy density of the bulk is given by the 5D cosmological constant. Here these assumptions cannot be rigorously true, but departures are usually small: even if there is exchange of energy between the bulk and the brane, in the situation of quasi equilibrium there is no net flux, while the energy density of the bulk is at least almost constant, because dominated by the 5D cosmological constant $\Lambda_5 \sim M_5^2$.

Along these lines one can now transfer to the warped solution the results which will be presented below for the flat scenario.

\section{Applications}\label{app}

As already mentioned, the constraints that can be extracted from the cosmology of KK particles gravitationally produced depend crucially on the detailed structure of the interaction such species have with ordinary fields. It is consequently mindful to analyse few cases explicitly. The first possibility to be taken into account is that of the gravitino field, simply because, once SUSY has been turned on, it is expected to necessarily possess a tower of KK modes, being part of the same multiplet the graviton is, therefore having access to the same spacetime dimensionality. One other possibility is gravitationally created axions, and will be discussed thereafter.

\subsection*{KK Gravitinos}

The gravitino is the fermionic superpartner of the graviton in SuperGravity (SUGRA) theories, which acquires a mass via the SUSY version of the Higgs mechanism. Its properties are uniquely defined by SUGRA, the only exception being the value of its mass, which depends on the particular mechanism that breaks SUGRA. Since this field is partnered with the graviton, once one expands the spacetime to five dimensions, it is natural to expect the gravitino to live in the fifth dimension as well, which in turn means that from a 4D observer it will appear as a tower of KK excitations.

In the case of the 4D gravitino, its lifetime is a function of its mass alone (as long as unstable gravitinos are concerned), being the coupling universally determined by the gravitational one, which also provides departure from thermal equilibrium throughout the expansion of the Cosmos. Hence, constraints on gravitino abundances in the early Universe are typically given for a certain mass range.

Light gravitinos, with mass $m_{3/2}$ below 100~GeV, are usually supposed to be the Lightest Supersymmetric Particles (LSP's) and therefore are surviving forever due to $\cal R$-parity conservation. Anyhow, even if this symmetry is badly broken, some of the lighter modes will still decay with a characteristic lifetime longer than the Universe's age. It is mandatory to ask then that such gravitinos do not overwhelm the Universe, in fact, they must not exceed the present day's DM energy density $\Omega_{\rm DM} h^2 \approx 0.12$.

If the gravitino mass is in the range 100~GeV -- 30~TeV, its decay can spoil the successful predictions of BBN: for small masses it catalyses the photodissociation of the light elements, while for large masses it injects pions into the plasma, altering the neutron to proton ratio at the onset of the BBN. Here the limits on gravitino abundance depend on the open decay channels and their branching ratios, but they usually are the most stringent ones. If the main decay mode is hadronic ($B_h \approx 1$), the constraints on the gravitino abundance $Y_{3/2}$ are roughly at the level $10^{-17} - 10^{-15}$, depending on the exact value of $m_{3/2}$. On the other hand, if the main decay mode is photon + neutralino, because the gravitino is lighter than the lightest coloured superparticle, $B_h$ is about $10^{-3}$ and the constraints on $Y_{3/2}$ are weaker, of order $10^{-16} - 10^{-12}$. For more details, see ref.~\cite{masahiro}.

The third option, although not easy to arrange in model building with soft SUSY breaking, is that the gravitino be much heavier than a few TeV. That implies a lifetime shorter than about 10$^{-2}$~s and no late time entropy release. However, if $\cal R$-parity is conserved, then for each gravitino that decays there will be an LSP at the end of its decay chain, which is then defined to be absolutely stable. Its abundance will reflect the initial gravitino one, up to $m_{3/2} \sim 10^5$~TeV (heavier gravitinos decay at $T \gtrsim 10$ GeV, which permits thermalisation of the generated LSP's), and must be kept under control in order to avoid conflict with the observed DM density. In this case, the bound on $Y_{3/2}$ is
\be
Y_{3/2} \le 6 \cdot 10^{-12} \, \left(\frac{100 \; {\rm GeV}}{m_{LSP}}\right) \, ,
\ee
where $m_{LSP}$ is the LSP mass, and is independent of the exact gravitino mass.

So far for the ordinary 4D gravitino, whose abundance produced by gravitational interaction was investigated in~\cite{noi1}. However, the picture can be dramatically mutated by the come into play of the KK states. In principle these states will have interaction properties other than the zero mode ones, being inherited from those of the 5D SUGRA model, most relevantly from the mechanism of SUSY breaking. The mass matrix for the KK states is in general non diagonal, leading to mass shifts possibly different state by state. For the sake of clarity though, and since the results outlined below are easily adapted to different models, each mode will be given the zero mode coupling, and the masses will be simply taken to be $m_n = m_{3/2} + n/R$. With the details of KK gravitinos interactions at hand, the bounds on their abundances are readily computed.

The first band corresponds to the lightest KK gravitinos, with a mass smaller than 100~GeV. Here we assume that the zero mode is much lighter than 100~GeV and thus $m_n \approx n/R$. Under this hypothesis, the number of lighter states is $N \simeq 9 \cdot 10^{28} / M_5^3$. The total abundance of light gravitinos today, and its relative constraint, would then be
\be\label{band1grav}
\Omega_{3/2} h^2 &\simeq& 8 \cdot 10^{32} M_5^{-21/4} \lesssim 0.12 \, ,\nonumber\\
\Rar M_5 &\gtrsim& 3 \cdot 10^6 \; \rm TeV \, .
\ee
Notice that this limit assumes stability for all the modes included in the summation. This may not be true if there are allowed transitions between different KK modes, not taken into account in this simplified exemplifying analysis. This is expected to be a good approximation for the light modes since these decays have typically narrower widths than the usual zeroth order gravitino~\cite{girma}, and are therefore expected to be more long lived.

The second interesting band encompasses KK modes with masses 100~GeV $\lesssim m_n \lesssim$ 30~TeV. Here the number of KK excitations is $N \simeq 3 \cdot 10^{31} / M_5^3$ and provides the following limit
\be\label{band2grav}
Y_{3/2} &\simeq& 8 \cdot 10^{30} M_5^{-21/4} \lesssim 10^{-12} \, ,\nonumber\\
\Rar M_5 &\gtrsim& 1.4 \cdot 10^8 \; \rm TeV \, .
\ee
This bound is actually relatively conservative, as within the band we have included also the more constraining 1~TeV gravitinos, which are typically (depending on their hadronic decay branching ratio) required to have $Y_{3/2} \lesssim 10^{-16}$. Indeed, if the main decay channels were hadronic one could pull the limit on $Y_{3/2}$ down to $10^{-14}$ which immediately pushes $M_5$ up to $3 \cdot 10^8 \; \rm TeV$ or higher. In any case, these limits are only illustrative, as there are more uncertainties affecting the precise value of $Y_{3/2}$, such as the more realistic numerical results for the number density of particles (which in the braneworld expansion regime is close to a factor of 10 more), and the possible presence of more thermalised relativistic degrees of freedom at that epoch.

Furthermore, here again there could be decays which violate the KK-number such as $KK^n \rar KK^m + X$, where $m < n$ and $X$ is a Standard Model (SM) particle. However their amplitudes is often negligible~\cite{girma} with respect to the standard decay channels (since also a transition $KK \rar LSP + X$ violates KK-number). This argument applies to the forthcoming paragraph as well.

The third and last band refers to the highest KK states. The upper bound can be taken to be of order $10^5$ TeV, leading to $N \simeq 9 \cdot 10^{34} / M_5^3$, and to a total $Y_{3/2}$ which is about $10^{11}$ times the one found in~(\ref{band2grav}), that in turn translates into a higher lower bound for $M_5$, that is,
\be\label{band3grav}
Y_{3/2} &\simeq& 1.6 \cdot 10^{42} M_5^{-21/4} \lesssim 10^{-12} \, ,\nonumber\\
\Rar M_5 &\gtrsim& 2 \cdot 10^{10} \; \rm TeV \, .
\ee
These limits are valid for $M_5 \lesssim 3 \cdot 10^{11} \; \rm TeV$ for in this case we have much more than 5 KK modes in the band.

So far it has always been assumed that the Universe, in the course of its evolution, has had available high enough temperatures, and fast enough expansion rates, for efficient gravitational particle production. This point is directly connected to the epoch which succeeded the inflationary expansion regime. This exponentially fast growing regime, besides providing another extremely well functioning source for gravitational creation of fields (although immediately exponentially diluted) is believed to have been followed by an abrupt phase transition which repopulated and thermalised the empty Universe, where again gravitational interaction could have played a major r\^ole. The temperature at the onset of the equilibrium era is usually calculated assuming instantaneous and efficient conversion of the energy stored in the field which drove the expansion, to that of the newly born SM (and its extensions) particles. Although not precise, this method gives an idea of the scales involved, and is readily worked out.

As a brief example, one can consider chaotic inflation scenarios on the brane. The reason for this choice, beyond its simplicity and its being favoured by observations, is that the completion of the reheating process happens in the brane expansion regime for most values of $M_5$. The potential for the scalar inflaton field is given by $V = m_\phi^2 \phi^2 /2$ and, as one learns from~\cite{roy,david}, successful generation of density perturbations in the brane regime requires $m_\phi \simeq 5 \cdot 10^{-5} M_5$, unlike the usual 4D models where the inflaton mass turns out to be close to the 4D gravity scale, as expected. If the scalar field interacts with fermions, its typical decay width can be parametrised by $\Gamma \simeq g_\phi m_\phi$, which gives a $\rho^2$ regime reheating temperature (obtained from equating $\Gamma$ to $H$)
\be\label{rehT}
T_R \simeq 0.05 g_\phi^{1/4} M_5 \gtrsim T_t \; \Rar \; M_5 \lesssim 2 \cdot 10^{12} g_\phi^{1/2} \, ,\nonumber
\ee
where the last step is required for self consistency. The highest mode available is therefore $N \simeq 5 \cdot 10^{25} g_\phi / M_5^2$, obtained by requiring that $H_R \simeq m_N$, and the overclosure constraints reads
\be\label{bandA3grav}
Y_{3/2} &\simeq& 9 \cdot 10^{11} g_\phi^{13/4} M_5^{-2} \lesssim 10^{-12} \, ,\nonumber\\
\Rar M_5 &\gtrsim& 10^{12} g_\phi^{13/8} \rm \; TeV \, .
\ee
Notice that this limit makes sense only if the modes produced do not weigh more than $m_{3/2} \simeq 10^5$~TeV, as is always the case for $g_\phi \lesssim 0.1$, for the same reasons explained above (above $g_\phi \simeq 0.1$ eq.~(\ref{band3grav}) applies). Lastly, the constraint~(\ref{band2grav}) coming from BBN is modified into
\be\label{bandA2grav}
M_5 \gtrsim 10^{13} g^{13/8} \; \rm TeV \, ,
\ee
using the hadronic bound, valid below $g_\phi \simeq 2 \cdot 10^{-3}$, and the lowest band limit~(\ref{band1grav}) becomes, up to $g_\phi \simeq 7 \cdot 10^{-4}$,
\be\label{bandA1grav}
M_5 \gtrsim 6 \cdot 10^{19} g^{17/4} \; \rm TeV \, .
\ee

\begin{figure}
\includegraphics[width=0.48\textwidth]{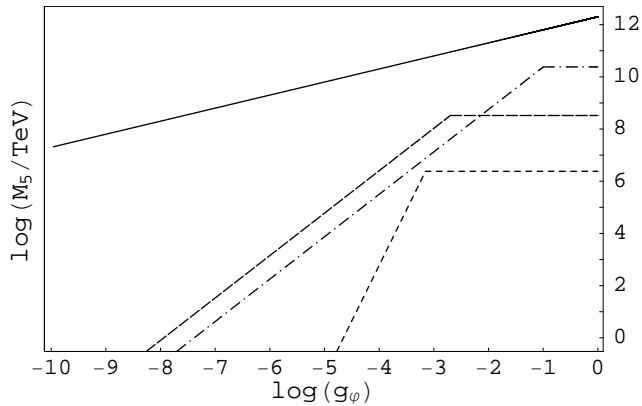}
\caption{\footnotesize{Limits on the 5D Planck mass $M_5$ plotted against the coupling constant $g_\phi$. The solid line is eq.~(\ref{rehT}), above it there is no $H \propto \rho$ regime after inflation, and no gravitational particle production. The lower dashed line is the constraint coming from the lowest KK modes: the region below the line is excluded (see eq.~(\ref{band1grav}) and its relative counterpart accounting for a low reheating temperature, eq.~(\ref{bandA1grav})). The upper dashed line corresponds to eqs.~(\ref{band2grav}) and (\ref{bandA2grav}), obtained by demanding sufficiently low entropy release during BBN. The region below the dot-dashed line is excluded instead thanks to the limits (\ref{band3grav}, \ref{bandA3grav}).}}
\label{plot}
\end{figure}

It is then possible to combine all the constraints in one single plot with the 5D Planck mass against the coupling constant $g_\phi$. By looking at the figure~\ref{plot} one sees how the lower bounds on $M_5$ are significantly weakened once a relatively low reheating temperature is taken into account, still leaving interesting constraints on the parameter(s) of the extra dimensions. When the inflaton coupling constant grows, the limits~(\ref{band1grav}), (\ref{band2grav}), and (\ref{band3grav}) are approached as expected. Other inflationary models are briefly mentioned in the appendix, where it is explained why chaotic inflation is probably the only relevant one to study in this context.

There is one more subtlety to be discussed here. As it has been shown in some SUSY realisations of flat or warped 5D models~\cite{susyX}, at the massive level (i.e.\ not for the zero mode) the typical amount of SUSY which is left after the compactification to four dimensions is higher then ${\cal N} = 1$. There is the possibility that more than one tower of KK gravitinos is present, and the interaction strengths of gravitinos belonging to different towers can be different. That in turn means that, as shown in~\cite{noi2}, $g_* \simeq g_*(MSSM) + (g_{3/2}+g_2) T/\Delta m$, which is even more favourable for gravitational particle production. Indeed higher energy density corresponding to the same temperature brings down the freezing out temperature (or can produce higher masses for the same temperature, as the Universe expands faster), strengthening the results outlined so far.

\subsection*{KK Axions}

Axions are one more example of a weakly interacting particle that may endanger the successful contact between the standard cosmological model and observational data. In the standard 4D theory, the axion pseudoscalar field is introduced in order to solve the strong CP problem~\cite{pq}, that is, to dynamically adjust the QCD $\Theta$ parameter to zero as a result of spontaneous symmetry breaking.

The anomalous coupling of the axion to gluons, which breaks the Peccei-Quinn (PQ) symmetry is
\be\label{axionL4D}
\mathcal{L}_{anom} = \xi \frac{g^2}{32 \pi^2} \, \frac{a}{f_4} \, G^2 \, ,
\ee
where $g$ is the strong coupling constant, $f_4$ the 4D decay constant, $G^2 \equiv G^{\mu\nu}_a \tilde G_{\mu\nu}^a$, with $G_{\mu\nu}^a$ the gluonic field strength tensor and $\tilde G_{\mu\nu}^a$ its dual, $\xi$ a model dependent dimensionless quantity, and $a$ is the actual axion field. The relation between the zero temperature axion mass,
$m_a$, and the axion decay constant $f_4$, is
\be\label{zeroAXmass4D}
m_a = \frac{z^{1/2}}{1+z} \frac{f_\pi \, m_\pi}{f_4} \, .
\ee
Here $f_\pi = 92$~MeV and $m_\pi = 135$~MeV are respectively the pion decay constant and mass, while $z = m_u/m_d \approx 0.6$ is the up to down quark mass ratio.

The 4D theory is able to provide certain bounds on the axion decay constant using cosmological and astrophysical arguments, which depend on the parameters of the exact axion model, but the allowed range is essentially (see e.g.\ ref.~\cite{pdg})
\be\label{boundAXf4D}
10^8 \; {\rm GeV} \lesssim f_4 \lesssim 10^{12} \; {\rm GeV} \, .
\ee
These limits imply a very light axion field, much lighter than 1 eV, as can be seen directly by use of~(\ref{zeroAXmass4D}), which makes its production through gravity irrelevant.

Once more, the scenario is radically different if the axion field lived in 5D. A specific 5D axion model was pushed forward in~\cite{tony}, its warped counterpart is found in~\cite{warpedax}. In their model the anomalous interaction term, after dimensional reduction, reads
\be\label{axionL5D1}
\mathcal{L}_{anom} = \xi \frac{g^2}{32 \pi^2} \, \left( \sum_0^\infty \frac{r_n \, a_n}{f_4} \right) \, G^2 \, ,
\ee
where $r_0 = 1$ and $r_{n\neq0} = \sqrt{2}$, and $a_n$ are the axion KK modes. These states are not the true mass eigenstates (see the discussion in~\cite{tony}), which are needed for the scope pursued in this paper, but it can be easily shown that for high $n$ these $a_n$ come closer and closer to the actual mass KK eigenstates. Hence, each KK axion excitation couples to ordinary matter (gluons as well as photons) through the same 4D decay constant, whereas the 5D original axion field couples at a 5D PQ breaking scale $f_5$ as
\be\label{axionL5D0}
\mathcal{L}_{anom} = \xi \frac{g^2}{32 \pi^2} \, \frac{a}{f_5} \, G^2 \, ,
\ee
where the two PQ scales are related by $f_4^2 = 2\pi M_5 f_5^3$ in the case of just one extra dimension.

It is useful to divide the KK axion tower into two bands, the long-living modes and the unstable ones. The latter include axions which could decay during and after BBN. However, unlike the gravitino, in this case no constraints can be derived for higher masses (lifetime shorter than $10^{-2}$ seconds) just by looking at BBN and DM density, since there is no symmetry that protects the axion from decaying into photons or gluons, and there will consequently be no dangerous relics as long as the decay happens before the BBN. The decay rate will depend explicitly on the mass of the axion and on $f_4$. In order to make the analysis traceable, the 4D PQ scale will be fixed to $f_4 \simeq 10^{10} \; \rm GeV$. In this work we are interested in gravitational creation of particles, while other mechanisms as well as limits obtainable from astrophysical arguments are discussed in~\cite{noi3}.

The first band is that for which the axions, ignoring possible transitions between different KK modes, are stable on the Universe's age timescale. The ($n$-th) axion decay rate $a \to 2\gamma$ is
\be\label{decayAX}
\Gamma_{2\gamma} &=& \frac{\alpha^2}{256 \pi^3} \, C \, \frac{m_a^3}{f_4^2} \, ,
\ee
where $\alpha$ is the fine structure constant and $C$ is a model dependent numerical coefficient close to 1. Since the channel has a branching ratio of order 1, for $C=1$ le lifetime is
\be
\tau = 2 \cdot 10^{10} \left(\frac{25 \; {\rm keV}}{m_a}\right)^3 \left(\frac{f_4}{10^{10} \; {\rm GeV}}\right)^2 \; {\rm yr}
\ee
and KK states with masses lower than about 25 keV have a lifetime exceeding the age of the Universe. The total number of these states is $N \simeq 2 \cdot 10^{22} / M_5^3$. In complete analogy with the gravitinos, their contribution to the density fraction has to be set below the DM one, which means that
\be\label{band1ax}
\Omega_a h^2 &\simeq& 4 \cdot 10^4 M_5^{-21/4} \lesssim 0.12 \, ,\nonumber\\
\Rar M_5 &\gtrsim& 11 \; \rm TeV \, .
\ee
This result is not very stringent because, despite their weak interactions, only the very light axions are able to survive for so long, which translates in a poorly relevant gravitationally produced relic density. This is especially true when compared to the bounds extracted from gravitinos, for they (likely) possess $\cal R$-symmetry, whose analogue is absent in the 5D axion scenario.

A more interesting bound is that derived from the second band, which includes axions whose decay would alter the primordial abundances of light elements. Having fixed the 4D PQ scale, the masses associated to a given lifetime are automatically fixed as well~(\ref{decayAX}).  

Now, recalling that $f_4 = 10^7 \; \rm TeV$, the states corresponding to a lifetime of about a second have masses around 20 GeV. These axions are strongly constraining if the decay branching ratio into hadrons is close to 1~\cite{kawasaki}, in which case they would efficiently alter the neutron to proton ratio during the early stages of BBN, affecting the final $^4$He abundance. These qualitative considerations, once put into numbers, lead to $M_5 \gtrsim 1.8 \cdot 10^5 \; \rm TeV$.

If the hadronic channel is instead suppressed, then the main effect is the photodissociation of light elements at times $t \sim 10^7 \; \rm s$ by decay of axions with mass about 100~MeV. The aforementioned potential problem with such axions can be avoided if the 5D Planck mass is pushed beyond $2 \cdot 10^5$~TeV.

Notice that in both cases the corresponding mass gaps are tiny on the scales of interest (0.1 to 20 GeV), thereby justifying the approximations adopted in the preceding discussion.

This is the most interesting limit coming from gravitational particle production of a 5D axion alone. Lastly, since the 4D PQ scale has been fixed, this lower limit on $M_5$ leads to a lower limit on the 5D axion decay constant, namely $f_5 \gtrsim 10^4 \; \rm GeV$.

One more comment is in order here. Since KK axions have weak but not gravitationally weak interactions, it is natural to ask whether they are in thermal equilibrium and when, for if that were the case then the gravitational abundances will be triggered to the usual thermal ones, erasing any information about their original production epoch. Consider for instance the 20 GeV KK excitation and the dominant $a \rar 2\gamma$ process. Roughly speaking, by comparing the rates of decay $\Gamma_{2\gamma}$ and of expansion $H$ it is possible to obtain an approximate freezing out temperature at which this particular interaction becomes highly improbable. In the case under investigation one sees that when $M_5 \gtrsim 1.3 \cdot 10^{11} \; \rm TeV$, then thermal equilibrium is established for $T_t \gtrsim T_{eq} \gtrsim 1.4 \cdot 10^{-25} M_5^3$, i.e.\ during the $H\sim\rho$ regime. Were the 5D Planck mass smaller, the inequality $\Gamma_{2\gamma} \lesssim H$ would return (this time in standard expansion regime) $T_{eq} \gtrsim 1.7 \cdot 10^9 \; {\rm TeV}$, but this temperature is higher than $T_t$ calculated for $M_5 \lesssim 1.3 \cdot 10^{11} \; \rm TeV$: there is never thermal equilibrium below the transition temperature. The constraints derived above are therefore valid, the freezing out temperature for gravitational interaction being smaller than that for which thermal equilibrium would be realised.

\section{Conclusion}

In this paper we have considered the phenomenology of gravitational particle production in the context of braneworld cosmology, where an epoch of non standard expansion ($H\sim\rho$) renders the mechanism appetible. Based on the results of~\cite{noi1}, we have extended the analysis to bulk fields that appear to a 4D observer as KK towers. As expected, the potential dangerousness of the produced abundances is able to strongly constrain the parameters of the extra dimensional model, confirming the well known fact that cosmology tends to dislike very low (e.g.\ TeV) gravity scales.

Since the crucial ingredient which makes gravitational particle production a cosmologically important mechanism is a period of braneworld expansion, this mechanism is intrinsically entwined with the early Universe's inflationary era, for the temperature (or the Hubble parameter) at the end of the reheating process depends, although the details are not fully understood, on the dynamics of the inflaton field.

Once a specific model of inflation is chosen, and a reheating scale is computed, the abundances of gravitationally produced particles are automatically obtained, therefore providing some information on the parameters of inflation. This is so because there is a whole KK tower of particles accessible by gravitational production, and the reheating temperature resulting from inflation dictates the upper cutoff for the actual available modes. Hence, the results hitherto described have interesting implications in braneworld inflation model building.

Two particular cases have been analysed in details, gravitinos and axions. This specialisation is required since, even though the production mechanism is basically insensitive on the properties of the generated field (the Equivalence Principle), it is their interactions with SM fields that is essential in determining their late time history, and consequently the limits that can be inferred from it.

If the 5D model is supersymmetrised then the presence of a KK tower (or more than one) of gravitinos strongly constrains the 5D Planck mass, at the level of $10^{10}$~TeV. This is especially true when the reheating process is efficient, allowing for extremely fastly expanding radiation dominated FRW Universe, which is the perfect environment for gravitational particle production.

A 5D axion is also able to put interesting bounds on $M_5$ because, even though its mass (the gravity charge) is typically very small, and consequently not efficiently produced by gravitational interactions, the existence of KK axions of any mass makes it potentially relevant. The bounds obtainable by demanding that the gravitationally produced axions do not release too much entropy during the process of BBN, are around $M_5 \gtrsim 10^5$~TeV (slightly model dependent) which, although not as strong as the KK gravitino bounds, is a fine example of how TeV scale gravity in the minimal extension of the SM employing a 5D axion field can be ruled out by simple cosmological arguments.

In general, although these explicit limits have been worked out only for two specific models, a similar analysis can be performed for any beyond the SM bulk field with little extra effort, as long as these fields do not substantially modify the brane Friedman equation~\ref{friedman}.

Concluding, gravitational particle production has a high phenomenological impact on braneworld cosmology, whose study provides yet another independent way of constraining the parameters of the extra dimensional spacetime.

\begin{acknowledgments}
We wish to thank Sasha Dolgov for useful comments and suggestions. F.U. thanks the organisers of the 2nd Workshop on Modern Cosmology held in the Benasque Center of Science, Spain, where this work has been completed. C.B. is supported in part by NSF under grant PHY-0547794 and by DOE under contract DE-FG02-96ER41005.
\end{acknowledgments}

\appendix

\section{Models of Braneworld Inflation}
In this brief appendix the reasons why the focus has been chosen to be on chaotic inflation only are outlined. The main motivations are simply summarisable as, first, it is the one that (in the context of braneworld inflation) fits the WMAP constraints the best, and secondly, has the widest parameters space for which after reheating a period of brane expansion ($H \propto \rho$) is likely to have taken place.

\subsection{Quartic Inflation}
The potential is $V = \lambda \phi^4 /4$. The self coupling is constrained by inflation to be $\lambda \simeq 8 \cdot 10^{-15}$~\cite{ed, david}. A possible mass term for the inflaton field must be subdominant with respect to the quartic term for the duration of inflation, in order for this to be pure quartic inflation (and for the limit just reported to apply), and then this implies that the decay rate (using the same parametrisation as in the body of the paper) is going to be much smaller than the standard chaotic one, thereby implying further less restrictive limits (on the mass of the field in this case). It is also likely that reheating completes during a standard $\rho$ dominated regime.

\subsection{Exponential Inflation}
The potential is $V = V_0 \exp^{-\alpha\phi / M_5}$, where the initial vacuum energy is encoded in $V_0$. This model has several problems, first of all in braneworld scenarios it needs to be supplemented with a curvaton field in order to obtain the correct spectrum of the perturbations (in particular the scalar-to-tensor ratio)~\cite{david}. Secondly it is not interesting as far as gravitational particle production is concerned because its reheating (that takes place through gravitational decay of the inflaton) is very inefficient and almost always occurs in standard cosmological expansion era~\cite{ed}.

\subsection{Natural Inflation}
The potential is $V = \Lambda^2 \left[ 1 + \cos(\phi/f) \right]$, where $f$ is the scale associated with the pseudo Nambu Goldstone boson $\phi$ and $\sqrt\Lambda$ is the scale of inflation (vacuum energy). One can obtain successful inflation on a brane with this potential provided that the following requirements are fulfilled~\cite{ed}: $\Lambda \simeq 0.1 M_5^2$ and $f \simeq 3 \cdot 10^2 M_5$. The mass of the inflaton field is given by $m_\phi \simeq \Lambda/f \simeq 3 \cdot 10^{-4} M_5$ and is slightly heavier that the chaotic scenario. However in this case reheating should proceed through loop interactions whose rate is $\Gamma \simeq m^3 / f^2 \simeq 3 \cdot 10^{-16} M_5$. This means that everything goes as in the chaotic case, only with fixed coupling constant $g_\phi \simeq 10^{-11}$, but this in turn then means that both $\Omega_X h^2$ and $Y_X$ are both at least reasonably small (see for instance eq.~\ref{bandA3grav}), hence, harmless.

\subsection{F-term and D-term Hybrid Inflation}
These models have been shown to be not realisable in the context of brane inflation, due to the (in)famous $\eta$ problem~\cite{ed}, that is, it is not possible to have a potential that is flat enough to let the Universe inflate and expand as long as needed: inflation is very short lived (in fact, this is a general problem in the context of supersymmetric realisations of inflation, in particular arising from string theory~\cite{liam}).

\section{Anomalous particle creation}

Conformally invariant fields which live in a conformally flat spacetime (i.e.\ a spacetime conformal to Minkowski, like all the 2D spacetimes or the FRW Universes) are not gravitationally produced: indeed, one can always perform a conformal transformation and obtain field equations which look like those in Minkowski spacetime, allowing to define unambiguously a vacuum state. Examples of conformally invariant fields are massless fermions, massless and conformally coupled scalars and, but only in 4D, the electromagnetic field. The presence of a mass term breaks conformal invariance and is responsible for particle production. Nevertheless, this is rigorously true only at the classical level, since quantum effects may change this conclusion, as consequence of the trace anomaly. 

Indeed, after renormalization, the trace of the expectation value of the energy-momentum tensor still vanishes for free massless fields in flat spacetime, but in more general cases it is not so. The trace anomaly has been calculated for self-interacting fields in Minkowski spacetime: a well known example is the trace of a gauge field~\cite{anomaly1}
\be
\langle T_\mu^\mu \rangle_{\rm ren} = \frac{\beta(g)}{2g} \langle G_{\kappa\lambda}^a G^{\kappa\lambda}_a \rangle_{\rm ren} \, ,
\ee
where $\beta = \mu \, \partial g/\partial \mu$ is the $\beta$-function of the renormalized coupling $g(\mu)$, while $G_{\kappa\lambda}^a$ is the strength tensor of the gauge field. Such a result can be easily understood by noticing that loop diagrams of $\langle T_\mu^\mu \rangle$ diverge and the gauge invariant regularization is not conformally invariant. The only other case in which the trace anomaly has been computed rigorously is that of a free field in curved spacetimes. In 4D, one finds~\cite{anomaly2}
\be\label{trace2}
\langle T_\mu^\mu \rangle_{\rm ren} &=& \frac{1}{2880 \pi^2} \Big[a_1 C_{\alpha\beta\gamma\delta} C^{\alpha\beta\gamma\delta} + \nonumber \\ && + a_2 \left(R_{\alpha\beta} R^{\alpha\beta} - \frac{1}{3} R^2\right) + a_3 \Box R + a_4 R^2\Big] \, , \nonumber \\
\ee
where $a_i$ are numerical coefficients which depend on the spin of the field. On the other hand, there is no trace anomaly for the case of an odd number of spacetime dimensions, which means that conformally invariant free fields propagating in a conformally flat 5D bulk would be still traceless at the quantum level and cannot be produced by an expanding FRW Universe.

As discussed in ref.~\cite{dolgov}, the amplitude of particle production in a conformally flat gravitational field is proportional to the trace of the energy-momentum tensor and therefore the trace anomaly can make the production of massless particles possible even in conformally flat spacetimes. If we consider the case of an $SU(N)$ gauge field, we find that for any power law expansion of the scale factor, that is $a(t) \propto t^q$, the rate of particle production per unit time and volume is roughly $\beta^2(g)/t^4$. We can therefore expect that, in the very early Universe, gravitational particle production via trace anomaly could be more efficient than the one due to the mass term. For example, if we indicate with $t_R$ the time of the Universe after inflation, the energy density of massless bosons produced by the expanding Universe to the cosmological energy density ratio in braneworld cosmology is
\be
\rho / \rho_c \sim \frac{\beta^2(g)}{M_5^3 \, t^3_R} \, ,
\ee
which can be compared with the counterpart in standard cosmology~\cite{dolgov}
\be
\rho / \rho_c \sim \frac{\beta^2(g)}{M_4^2 \, t^2_R} \, .
\ee
For $M_4 \gg M_5$ the mechanism may be efficient. Nevertheless, interesting phenomenological consequences are unlikely: the process produces massless gauge bosons and, including radiative corrections, charged particles. On the other hand, interesting signatures would require weakly interacting particles which are never in thermal equilibrium after being produced gravitationally and whose decay during or after the BBN may leave observable effects. From this point of view, there are currently no theoretically well motivated candidates which can do the job. 

The non--zero trace in eq.~(\ref{trace2}) may instead be responsible for the gravitational production of sterile or very weakling interacting particles. However, it is typically a very small quantity and, even if at high temperature may be more important than the production via the mass term (e.g.\ for superlight particles), it is not likely that it can lead to interesting and observable effects.

\end{document}